\documentstyle{article}
\begin{document}
\begin{center}
\Large{\bf  ON GEOMETRY OF THE R\"OSSLER\\[2mm] SYSTEM OF
EQUATIONS }\vspace{4mm}\normalsize
\end{center}
 \begin{center}
\Large{\bf Valery Dryuma}\vspace{4mm}\normalsize
\end{center}
\begin{center}
{\bf Institute of Mathematics and Informatics AS Moldova, Kishinev}\vspace{4mm}\normalsize
\end{center}
\begin{center}
{\bf  E-mail: valery@dryuma.com;\quad cainar@mail.md}\vspace{4mm}\normalsize
\end{center}
\begin{center}
{\bf  Abstract}\vspace{4mm}\normalsize
\end{center}

     Theory of Riemann extensions of spaces of constant affine connection
     is proposed to study  the R\"ossler system
    of equations
\[
    \frac{dx}{ds}=-(y+z),\quad
  \frac{dy}{ds}=x+\alpha y,\quad \frac{dz}{ds}=\beta+xz-\nu z.
\]

      After its presentation in a homogeneous form
\[
{\frac
{d}{ds}}\xi(s)+1/5\,{\xi}^{2}-1/5\,\xi\,\rho\,\nu+1/5\,\xi\,
\rho\,\alpha+\eta\,\rho+\theta\,\rho=0
\]
\[
{\frac
{d}{ds}}\eta(s)+1/5\,\xi\,\eta-\xi\,\rho-1/5\,\eta\,\rho\,\nu-4
/5\,\eta\,\rho\,\alpha=0
\]
\[
{\frac
{d}{ds}}\theta(s)-4/5\,\xi\,\theta+4/5\,\theta\,\rho\,\nu+1/5\,
\theta\,\rho\,\alpha-\beta\,{\rho}^{2}=0
\]
\[
{\frac
{d}{ds}}\rho(s)+1/5\,\xi\,\rho-1/5\,{\rho}^{2}\nu+1/5\,{\rho}^{
2}\alpha=0,
\]
where
\[
x=\frac{\xi}{\rho},\quad y=\frac{\eta}{\rho},\quad
z=\frac{\theta}{\rho},
\]
it can be considered as geodesic equations
\[
    \frac{d^2 X^i}{d s^2} +\Pi^i_{jk}\frac{d X^j}{d s}\frac{d X^k}{d
    s}=0.
\]
of four dimensional space
$M^4$ of constant affine connection with the components
$\Pi^i_{jk}=\Pi^i_{jk}(\alpha,\beta,\nu)$ depending on
parameters.

 \section{ From the first order system of equations
to the second order systems of ODE}

   The systems of the first order differential equations
\begin{equation} \label{dryuma:eq1}
\frac{d x^{i}}{ds}=c^i+a^{i}_{j}x^{j}+b^{i}_{ j k}x^{j}x^{k}
\end{equation}
depending on the parameters $a,b,c$ are not suitable
   object of consideration from  usually point of the Riemann geometry.

   The systems of the second order differential equations in form
\begin{equation} \label{dryuma:eq6}
\frac{d^2 x^i}{ds^2}+\Pi^i_{k j}(x)\frac{d x^k}{ds}\frac{d
x^j}{ds}=0
\end{equation}
are best suited to do that.

   They can be considered as geodesic of  affinely connected space
   $M^k$ in local coordinates $x^k$. The values $\Pi^i_{jk}=\Pi^i_{kj}$ are the coefficients of
   affine connections on $M^k$.

    With the help of such coefficients can
   be constructed curvature tensor and others geometrical objects
   defined on variety $M^k$.

\section{From  affinely connected space to the
Riemann space}

     We shall construct the Riemann space starting from a
     given affinely connected space defined by the systems of the second order
     ODE's.

      With this aim we use the notion of the Riemann extension of nonriemannian  space which was used
      earlier in the articles of author.

     Remind the basic properties of this construction.

     With help of the coefficients of affine connection of a given n-dimensional space
      can be introduced  2n-dimensional
     Riemann space $D^{2n}$ in local coordinates $(x^i,\Psi_i)$ having the metric of form
\begin{equation} \label{dryuma:eq11}
{^{2n}}ds^2=-2\Pi^k_{ij}(x^l)\Psi_k dx^i dx^j+2d \Psi_k dx^k
\end{equation}
\noindent where $\Psi_{k}$ are an additional coordinates.

  The important property of such type metric is that the geodesic
 equations of metric (\ref{dryuma:eq11})  decomposes into two parts
\begin{equation} \label{dryuma:eq12}
\ddot x^k +\Pi^k_{ij}\dot x^i \dot x^j=0,
\end{equation}
and
\begin{equation} \label{dryuma:eq13}
\frac{\delta^2 \Psi_k}{ds^2}+R^l_{kji}\dot x^j \dot x^i \Psi_l=0,
\end{equation}
where
\[
\frac{\delta \Psi_k}{ds}=\frac{d
\Psi_k}{ds}-\Pi^l_{jk}\Psi_l\frac{d x^j}{ds}
\]
and $R^l_{kji}$ are the curvature tensor of n-dimensional space
with a given affine connection.

 The first part (\ref{dryuma:eq12}) of the full system
is the system of equations for geodesic of basic space with local
coordinates $x^i$ and it do not contains the supplementary
coordinates $\Psi_k$.

 The second part (\ref{dryuma:eq13}) of the system  has the form
of linear $N\times N$ matrix system of second order ODE's for
supplementary  coordinates $\Psi_k$
\begin{equation} \label{dryuma:eq14}
\frac{d^2 \vec \Psi}{ds^2}+A(s)\frac{d \vec \Psi}{ds}+B(s)\vec
\Psi=0.
\end{equation}

   Remark that the full system of geodesics has the first integral
\begin{equation} \label{dryuma:eq15}
-2\Pi^k_{ij}(x^l)\Psi_k \frac{dx^i}{ds}\frac{dx^j}{ds}+2\frac{d
\Psi_k}{ds}\frac{dx^k}{ds}=\nu
\end{equation}
which is equivalent to the relation
\begin{equation} \label{dryuma:eq16}
2\Psi_k\frac{dx^k}{ds}=\nu s+\mu
\end{equation}
where $\mu, \nu$ are parameters.

   The geometry of extended space
connects with geometry of basic space. For example the property of
the space to be Ricci-flat $R_{ij}=0$ or symmetrical
$R_{ijkl;m}=0$ keeps also for an extended space.

    It is important to note that for extended  space having the metric (\ref{dryuma:eq11})
    all scalar curvature invariants are vanished.

    As consequence the properties of linear system of
equation (\ref{dryuma:eq13}-\ref{dryuma:eq14}) depending from the
the invariants of  $N\times N$ matrix-function
\[
E=B-\frac{1}{2}\frac{d A}{ds}-\frac{1}{4}A^2
\]
under change of the coordinates $\Psi_k$ can be of used for that.

   First applications the notion of extended spaces for the studying
of nonlinear second order ODE's connected with nonlinear dynamical
systems have been considered by author (V.Dryuma 2000-2008).

\section{ Eight-dimensional Riemann space for the R\"ossler
system of equations}

     To investigation  the properties of the R\"ossler system equations
 \begin{equation}\label{dryuma:eq19}
    \frac{dx}{ds}=-(y+z),\quad
  \frac{dy}{ds}=x+\alpha y,\quad \frac{dz}{ds}=\beta+xz-\nu z
\end{equation}
we use its presentation in homogeneous form
\[
{\frac
{d}{ds}}\xi(s)+1/5\,{\xi}^{2}-1/5\,\xi\,\rho\,\nu+1/5\,\xi\,
\rho\,\alpha+\eta\,\rho+\theta\,\rho=0
\]
\[
{\frac
{d}{ds}}\eta(s)+1/5\,\xi\,\eta-\xi\,\rho-1/5\,\eta\,\rho\,\nu-4
/5\,\eta\,\rho\,\alpha=0
\]
\[
{\frac
{d}{ds}}\theta(s)-4/5\,\xi\,\theta+4/5\,\theta\,\rho\,\nu+1/5\,
\theta\,\rho\,\alpha-\beta\,{\rho}^{2}=0
\]
\[
{\frac
{d}{ds}}\rho(s)+1/5\,\xi\,\rho-1/5\,{\rho}^{2}\nu+1/5\,{\rho}^{
2}\alpha=0,
\]
where
\[
x=\frac{\xi}{\rho},\quad y=\frac{\eta}{\rho},\quad
z=\frac{\theta}{\rho}.
\]

   The relation between both systems is defined by the conditions
\[
   x(s)=\frac{\xi}{\rho},\quad y(s)=\frac{\eta}{\rho},\quad
   z(s)=\frac{\theta}{\rho}.
\]

    Remark that for a given system
\[\dot \xi=P,\quad
\dot \eta=Q,\quad \dot \theta=R,\quad \dot \rho=T
\] the condition
\[
P_{\xi}+Q_{\eta}+R_{\theta}+T_{\rho}=0
\]
is fulfield.

    Such type of the system  can be rewriten in the form
\[
    \frac{d^2 X^i}{d s^2} +\Pi^i_{jk}\frac{d X^j}{d s}\frac{d X^k}{d
    s}=0,
\]
which allow us to consider it as geodesic equations of the space
with constant  affine connection.

    In our case nonzero components of connection are
    \[
    \Pi^1_{11}=\frac{1}{5},\quad
    \Pi^1_{14}=\frac{\alpha-\nu}{10},\quad
    \Pi^1_{24}=\frac{1}{2},\quad  \Pi^1_{34}=\frac{1}{2},
    \quad \Pi^2_{12}=\frac{1}{10},\]\[ \Pi^2_{14}=-\frac{1}{2},\quad
    \Pi^2_{24}=-\frac{4 \alpha+\nu}{10},\quad \Pi^3_{34}=\frac{4 \nu+\alpha}{10},
    \]\[ \Pi^3_{12}=-\frac{4}{10},\quad \Pi^3_{44}=-\beta,
\quad \Pi^4_{14}=\frac{1}{10},\quad
\Pi^4_{44}=\frac{\alpha-\nu}{5}.
    \]

     The metric of corresponding Riemann space is
      \begin{equation}\label{dryuma:eq20}
     ^{8}ds^2=-2\, \Pi^1_{11}\,P dx^{2}+2\,\left
(-2\,\Pi^2_{12}\,Q-2\,\Pi^3_{12}\,U\right ) dx dy+2\,\left
(-2\,\Pi^1_{14}\,P-2\,\Pi^2_{14}\,Q-2\, \Pi^4_{14}\,V\right
)dxdu+\]\[+2\,\left (-2\,\Pi^1_{24}\,P-2\,\Pi^2_{24}\,Q\right ) dy
du+2\,\left (-2 \,\Pi^1_{34}\,P-2\, \Pi^3_{34}\,U\right )dz
du+\]\[+\left (-2\,\Pi^3_{44}\,U-2\, \Pi^4_{44}\,V\right )
du^{2}+2\,dx dP+2\, dydQ+2\,dz dU+2\,du dV ,
\end{equation}
where $(P,Q,U,V)$ are an additional coordinates.

    Geodesic of the metric (\ref{dryuma:eq20}) for coordinates $(x,y,z,u)$ are
\[
{\frac {d^{2}}{d{s}^{2}}}x(s)+1/5\,\left ({\frac
{d}{ds}}x(s)\right )^ {2}+1/5\,\left ({\frac {d}{ds}}u(s)\right
)\left ({\frac {d}{ds}}x(s) \right )\alpha-1/5\,\left ({\frac
{d}{ds}}u(s)\right )\left ({\frac {d }{ds}}x(s)\right )\nu+\left
({\frac {d}{ds}}u(s)\right ){\frac {d}{ds} }y(s)+\]\[+\left
({\frac {d}{ds}}u(s)\right ){\frac {d}{ds}}z(s)=0,
\]
\[
{\frac {d^{2}}{d{s}^{2}}}y(s)+1/5\,\left ({\frac
{d}{ds}}y(s)\right ){ \frac {d}{ds}}x(s)-\left ({\frac
{d}{ds}}u(s)\right ){\frac {d}{ds}}x( s)-4/5\,\left ({\frac
{d}{ds}}u(s)\right )\left ({\frac {d}{ds}}y(s) \right
)\alpha-1/5\,\left ({\frac {d}{ds}}u(s)\right )\left ({\frac {d
}{ds}}y(s)\right )\nu  =0,
\]
\[
{\frac {d^{2}}{d{s}^{2}}}z(s)-4/5\,\left ({\frac
{d}{ds}}y(s)\right ){ \frac {d}{ds}}x(s)+4/5\,\left ({\frac
{d}{ds}}u(s)\right )\left ({ \frac {d}{ds}}z(s)\right
)\nu+1/5\,\left ({\frac {d}{ds}}u(s)\right ) \left ({\frac
{d}{ds}}z(s)\right )\alpha-\beta\,\left ({\frac {d}{ds}}
u(s)\right )^{2} =0,
\]
\[
{\frac {d^{2}}{d{s}^{2}}}u(s)+1/5\,\left ({\frac
{d}{ds}}u(s)\right ){ \frac {d}{ds}}x(s)+1/5\,\left ({\frac
{d}{ds}}u(s)\right )^{2}\alpha-1 /5\,\left ({\frac
{d}{ds}}u(s)\right )^{2}\nu=0
\]
and  they have a  form of homogeneous R\"ossler system in the
variables
\[\xi=\frac{d x}{d s},\quad \eta=\frac{d y}{d s},\theta=\frac{d z}{d s},\quad
\rho=\frac{d u}{d s}.
\]

   The system of second order differential equations for additional coordinates
 can be reduced to the linear system of the first order equations
 with variable coefficients
     \[
{\frac {d}{dt}}P(t)+A_1 P(t)+B_1 Q(t)+C_1U(t)+E_1V(t) =0,
\]
\[
{\frac {d}{dt}}Q(t)+A_2P(t)+B_2Q(t)+C_2U(t)+E_2V(t) =0,
\]
\[
{\frac {d}{dt}}U(t)+A_3P(t)+B_3Q(t)+C_3U(t)+E_3V(t) =0,
\]
\[
{\frac {d}{dt}}V(t)+A_4P(t)+B_4Q(t)+C_4U(t)+E_4V(t) =0,
\]
where $ A_i,~B_i,~C_i,~E_i$ are the functions of the variables
$x,~y,~z,~u)$.

     Properties of such type of the systems can be investigated with help of the
     Wilczynski invariants.

\section{Laplace operator}

   In theory of Riemann spaces the equation
 \begin{equation}\label{dryuma:eq21}
L\psi=g^{ij}(\frac{\partial ^2}{\partial x^i \partial x
^j}-\Gamma^k_{ij}\frac{\partial}{\partial x^k})\psi(x)=0
\end{equation}
can be used to the study of the properties of spaces.

For the eight-dimensional space with the metric (\ref{dryuma:eq20})
corresponded the R\"ossler system we get the equation on the
function $\psi(x,y,z,u,P,Q,U,V)=\theta(P,Q,U,V)$
\[
4/5\,{\frac {\partial }{\partial P}}\theta(P,Q,U,V)+2/5\,P{\frac {
\partial ^{2}}{\partial {P}^{2}}}\theta(P,Q,U,V)-8/5\,\left ({\frac {
\partial ^{2}}{\partial P\partial Q}}\theta(P,Q,U,V)\right )U+\]\[+2/5\,
\left ({\frac {\partial ^{2}}{\partial P\partial
Q}}\theta(P,Q,U,V) \right )Q-2\,\left ({\frac {\partial
^{2}}{\partial P\partial V}} \theta(P,Q,U,V)\right )Q+2/5\,\left
({\frac {\partial ^{2}}{\partial P
\partial V}}\theta(P,Q,U,V)\right )V+\]\[+2\,\left ({\frac {\partial ^{2}}{
\partial Q\partial V}}\theta(P,Q,U,V)\right )P+2\,\left ({\frac {
\partial ^{2}}{\partial U\partial V}}\theta(P,Q,U,V)\right )P+8/5\,
\left ({\frac {\partial }{\partial V}}\theta(P,Q,U,V)\right
)\alpha+\]\[+2/ 5\,\left ({\frac {\partial }{\partial
V}}\theta(P,Q,U,V)\right )\nu-2/ 5\,\left ({\frac {\partial
^{2}}{\partial {V}^{2}}}\theta(P,Q,U,V) \right )V\nu+2/5\,\left
({\frac {\partial ^{2}}{\partial {V}^{2}}} \theta(P,Q,U,V)\right
)V\alpha+\]\[+2/5\,\left ({\frac {\partial ^{2}}{
\partial U\partial V}}\theta(P,Q,U,V)\right )U\alpha+2/5\,\left ({
\frac {\partial ^{2}}{\partial Q\partial V}}\theta(P,Q,U,V)\right
)Q \nu+8/5\,\left ({\frac {\partial ^{2}}{\partial Q\partial
V}}\theta(P, Q,U,V)\right )Q\alpha-\]\[-2/5\,\left ({\frac
{\partial ^{2}}{\partial P
\partial V}}\theta(P,Q,U,V)\right )P\nu-2\,\left ({\frac {\partial ^{2
}}{\partial {V}^{2}}}\theta(P,Q,U,V)\right )\beta\,U+8/5\,\left ({
\frac {\partial ^{2}}{\partial U\partial V}}\theta(P,Q,U,V)\right
)U \nu+\]\[+2/5\,\left ({\frac {\partial ^{2}}{\partial P\partial
V}}\theta(P, Q,U,V)\right )P\alpha =0.
\]

   This equation has varies type of particular solutions.

   A  simplest one is
\[
\psi(x,y,z,u,P,Q,U,V)={e^{\left (\nu-\alpha\right )P-5\,Q+5\,U+V}}
\]
at the conditions on parameters of the R\"ossler system
\[
\beta={\frac {25}{2}}\,\nu,\quad \alpha=-3/2\,\nu.
\]

    As examples obtained by direct substitutions we get  the
    quadratic solution
    \[
    \theta(P,Q,U,V)=\]\[=1/18\,{\frac {9\,{Q}^{2}+30\,QP\alpha-18\,QV+25\,{P}^{
2}{\alpha}^{2}-30\,P\alpha\,V+9\,{V}^{2}+15\,{P}^{2}+60\,U\alpha\,P+45
\,\beta\,UP}{{\it \_c}_{{2}}}}
\]
with conditions
\[
\nu=8/3\,\alpha,\quad \beta=arbitrary.
\]
    Cubic solution
    \[
    \theta(P,Q,U,V)=-{\frac {1}{240}}\,{\frac {\left (-96\,\alpha\,{\it l3
}-2880\,{\alpha}^{2}{\it l1}\right ){V}^{3}}{{\beta}^{2}}}-{\frac
{1}{ 240}}\,{\frac {\left (-48\,\beta\,{\it l3}-1440\,\beta\,{\it
l1}\, \alpha\right )U{V}^{2}}{{\beta}^{2}}}+\]\[+\left (-{\frac
{1}{240}}\,{ \frac {\left (60\,\beta\,{\it l3}+1800\,\beta\,{\it
l1}\,\alpha\right ){P}^{2}}{{\beta}^{2}}}+{\it l1}\,{U}^{2}\right
)V+{\it k1}\,{U}^{3}+ \left ({\it l3}\,P+{\it l2}\,Q\right
){U}^{2}-\]\[-{\frac {1}{240}}\,{ \frac {\left (2250\,\beta\,{\it
l1}\,\alpha+75\,\beta\,{\it l3}\right
){P}^{2}Q}{{\beta}^{2}}}+\left ({\it m2}\,{Q}^{2}-{\frac
{1}{240}}\,{ \frac {\left (400\,{\beta}^{2}{\it
l1}+3000\,\beta\,{\it l1}\,\alpha+ 100\,\beta\,{\it l3}\right
){P}^{2}}{{\beta}^{2}}}\right )U+n{Q}^{3}
\]
at the condition
\[
\nu=6 \alpha
\]
and arbitrary coefficients ${\it l},{\it m},{\it k},{\it n}$.

     A polynomial  solution of degree four
\[
\theta(P,Q,U,V)=r{Q}^{4}+{\it k1}\,{U}^{4}-2/5\,{\it
m3}\,{U}^{3}V+{ \it l2}\,U{Q}^{3}+{\it m2}\,{U}^{2}{Q}^{2}+{\it
n2}\,{U}^{3}Q+{\it m3} \,{U}^{2}{P}^{2}
\]
at the condition
\[
\nu=-{\frac {7}{13}}\,\alpha.
\]

     Remark that the properties of of such type of solutions depend on parameters and may be
     highly diversified.

 More complicated solutions of the Laplace equation can be obtained by application of the  method of
    $(u,v)$- transformation developed in the works of author.

    \section{Eikonal equation}

    Solutions of eikonal equation
 \begin{equation}\label{dryuma:eq22}
g{{^i}{^j}}\frac{\partial F}{\partial x^i}\frac{\partial
F}{\partial x^j}=0 \end{equation}
 also gives useful information
about the properties of Riemann space.

   In particular the condition
   \[
   F(x^1,x^2,...,x^i)=0
   \]
  where function $F(x^i)$ satisfies the equation $(\ref{dryuma:eq22})$,
   determines $(N-1)$-dimensional hypersurface with  normals  forming an isotropic
   vector field.

    For the space with the metric (\ref{dryuma:eq20})  the
    eikonal equation on the function
    $\psi(x,y,z,u,P,Q,U,V)=\eta(P,Q,U,V)$  takes the form
     \begin{equation}\label{dryuma:eq23}
    2/5\,P\left ({\frac {\partial }{\partial P}}\eta(P,Q,U,V)\right )^{2}-
8/5\,\left ({\frac {\partial }{\partial P}}\eta(P,Q,U,V)\right
)\left ({\frac {\partial }{\partial Q}}\eta(P,Q,U,V)\right
)U+\]\[+2/5\,\left ({ \frac {\partial }{\partial
P}}\eta(P,Q,U,V)\right )\left ({\frac {
\partial }{\partial Q}}\eta(P,Q,U,V)\right )Q-2\,\left ({\frac {
\partial }{\partial P}}\eta(P,Q,U,V)\right )\left ({\frac {\partial }{
\partial V}}\eta(P,Q,U,V)\right )Q+\]\[+2/5\,\left ({\frac {\partial }{
\partial P}}\eta(P,Q,U,V)\right )\left ({\frac {\partial }{\partial V}
}\eta(P,Q,U,V)\right )P\alpha-2/5\,\left ({\frac {\partial
}{\partial P}}\eta(P,Q,U,V)\right )\left ({\frac {\partial
}{\partial V}}\eta(P,Q ,U,V)\right )P\nu+\]\[+2/5\,\left ({\frac
{\partial }{\partial P}}\eta(P,Q, U,V)\right )\left ({\frac
{\partial }{\partial V}}\eta(P,Q,U,V)\right )V+2\,\left ({\frac
{\partial }{\partial Q}}\eta(P,Q,U,V)\right ) \left ({\frac
{\partial }{\partial V}}\eta(P,Q,U,V)\right )P-\]\[-8/5\, \left
({\frac {\partial }{\partial Q}}\eta(P,Q,U,V)\right )\left ({
\frac {\partial }{\partial V}}\eta(P,Q,U,V)\right
)Q\alpha-2/5\,\left ({\frac {\partial }{\partial
Q}}\eta(P,Q,U,V)\right )\left ({\frac {
\partial }{\partial V}}\eta(P,Q,U,V)\right )Q\nu+\]\[+2\,\left ({\frac {
\partial }{\partial U}}\eta(P,Q,U,V)\right )\left ({\frac {\partial }{
\partial V}}\eta(P,Q,U,V)\right )P+8/5\,\left ({\frac {\partial }{
\partial U}}\eta(P,Q,U,V)\right )\left ({\frac {\partial }{\partial V}
}\eta(P,Q,U,V)\right )U\nu+\]\[+2/5\,\left ({\frac {\partial
}{\partial U}} \eta(P,Q,U,V)\right )\left ({\frac {\partial
}{\partial V}}\eta(P,Q,U, V)\right )U\alpha+2/5\,\left ({\frac
{\partial }{\partial V}}\eta(P,Q, U,V)\right
)^{2}V\alpha-\]\[-2/5\,\left ({\frac {\partial }{\partial V}}
\eta(P,Q,U,V)\right )^{2}V\nu-2\,\left ({\frac {\partial
}{\partial V} }\eta(P,Q,U,V)\right )^{2}\beta\,U =0.
\end{equation}

    A simplest solution of this equation is
    \[
\eta(P,Q,U,V)=-{\frac {Q}{\alpha}}+{\frac {V}{\nu-\alpha}}+{\frac
{U}{ \alpha}}+P
\]
with condition on the coefficients of the R\"ossler system
  \begin{equation}\label{dryuma:eq24}
-5\,\beta\,\alpha-11\,\alpha\,\nu+8\,{\nu}^{2}+3\,{\alpha}^{2} =0.
\end{equation}
    From here we find
    \[
\nu={\frac {11}{16}}\,\alpha+1/16\,\sqrt
{25\,{\alpha}^{2}+160\,\beta\, \alpha}
\]

   To provide a more complicated solutions of the equation
$(\ref{dryuma:eq22})$ we use the method of $(u,v)$-transformation.

    For the sake of convenience we rewrite the equation
    $(\ref{dryuma:eq22})$ in the form
     \begin{equation}\label{dryuma:eq25}
    2/5\,x\left ({\frac {\partial }{\partial x}}\eta(x,y,z,p)\right )^{2}-
8/5\,\left ({\frac {\partial }{\partial x}}\eta(x,y,z,p)\right
)\left ({\frac {\partial }{\partial y}}\eta(x,y,z,p)\right
)z+\]\[+2/5\,\left ({ \frac {\partial }{\partial
x}}\eta(x,y,z,p)\right )\left ({\frac {
\partial }{\partial y}}\eta(x,y,z,p)\right )y-2\,\left ({\frac {
\partial }{\partial x}}\eta(x,y,z,p)\right )\left ({\frac {\partial }{
\partial p}}\eta(x,y,z,p)\right )y+$$$$+2/5\,\left ({\frac {\partial }{
\partial x}}\eta(x,y,z,p)\right )\left ({\frac {\partial }{\partial p}
}\eta(x,y,z,p)\right )x\alpha-2/5\,\left ({\frac {\partial
}{\partial x}}\eta(x,y,z,p)\right )\left ({\frac {\partial
}{\partial p}}\eta(x,y ,z,p)\right )x\nu+$$$$+2/5\,\left ({\frac
{\partial }{\partial x}}\eta(x,y, z,p)\right )\left ({\frac
{\partial }{\partial p}}\eta(x,y,z,p)\right )p+2\,\left ({\frac
{\partial }{\partial y}}\eta(x,y,z,p)\right ) \left ({\frac
{\partial }{\partial p}}\eta(x,y,z,p)\right )x-$$$$-8/5\, \left
({\frac {\partial }{\partial y}}\eta(x,y,z,p)\right )\left ({
\frac {\partial }{\partial p}}\eta(x,y,z,p)\right
)y\alpha-2/5\,\left ({\frac {\partial }{\partial
y}}\eta(x,y,z,p)\right )\left ({\frac {
\partial }{\partial p}}\eta(x,y,z,p)\right )y\nu+$$$$+2\,\left ({\frac {
\partial }{\partial z}}\eta(x,y,z,p)\right )\left ({\frac {\partial }{
\partial p}}\eta(x,y,z,p)\right )x+8/5\,\left ({\frac {\partial }{
\partial z}}\eta(x,y,z,p)\right )\left ({\frac {\partial }{\partial p}
}\eta(x,y,z,p)\right )z\nu+$$$$+2/5\,\left ({\frac {\partial
}{\partial z}} \eta(x,y,z,p)\right )\left ({\frac {\partial
}{\partial p}}\eta(x,y,z, p)\right )z\alpha+2/5\,\left ({\frac
{\partial }{\partial p}}\eta(x,y, z,p)\right
)^{2}p\alpha-2/5\,\left ({\frac {\partial }{\partial p}}
\eta(x,y,z,p)\right )^{2}p\nu-$$$$-2\,\left ({\frac {\partial
}{\partial p} }\eta(x,y,z,p)\right )^{2}\beta\,z=0
\end{equation}

     Now after change of the function and derivatives in accordance with the rules
\[
\eta(x,y,z,p) \rightarrow u(x,t,z,p),\quad y \rightarrow
v(x,t,z,p,
\]
     \[
\frac{\partial \eta(x,y,z,p)}{\partial x} \rightarrow
\frac{\partial u(x,t,z,p)}{\partial x}-\frac{\frac{\partial
u(x,t,z,p)}{\partial t}}{\frac{\partial v(x,t,z,p))}{\partial
t}}\frac{\partial v(x,t,z,p)}{\partial x},
\]
\[
\frac{\partial \eta(x,y,z,p)}{\partial z} \rightarrow
\frac{\partial u(x,t,z,p)}{\partial z}-\frac{\frac{\partial
u(x,t,z,p)}{\partial t}}{\frac{\partial v(x,t,z,p))}{\partial
t}}\frac{\partial v(x,t,z,p)}{\partial z},
\]
\[
\frac{\partial \eta(x,y,z,p)}{\partial p} \rightarrow
\frac{\partial u(x,t,z,p)}{\partial p}-\frac{\frac{\partial
u(x,t,z,p)}{\partial t}}{\frac{\partial v(x,t,z,p))}{\partial
t}}\frac{\partial v(x,t,z,p)}{\partial p},
\]
\[
\frac{\partial \eta(x,y,z,p)}{\partial y} \rightarrow
\frac{\frac{\partial u(x,t,z,p)}{\partial t}}{\frac{\partial
v(x,t,z,p))}{\partial t}},
\]
where
\[u(x,t,z,p)=t{\frac {\partial }{\partial t}}\omega(x,t,z,p)-\omega(x,t,
z,p),\quad v(x,t,z,p)={\frac {\partial }{\partial
t}}\omega(x,t,z,p)
\]
we find the equation on the  function $\omega(x,t,z,p)$
 \begin{equation}\label{dryuma:eq26} \left
({\frac {\partial }{\partial x}}\omega(x,t,z,p)\right )^{2}x-5\,
tx{\frac {\partial }{\partial p}}\omega(x,t,z,p)-p\nu\,\left
({\frac {
\partial }{\partial p}}\omega(x,t,z,p)\right )^{2}-5\,\beta\,z\left ({
\frac {\partial }{\partial p}}\omega(x,t,z,p)\right )^{2}+
$$$$+p\left ({ \frac {\partial }{\partial x}}\omega(x,t,z,p)\right
){\frac {\partial }{\partial p}}\omega(x,t,z,p)+4\,z\nu\,\left
({\frac {\partial }{
\partial z}}\omega(x,t,z,p)\right ){\frac {\partial }{\partial p}}
\omega(x,t,z,p)+4\,\left ({\frac {\partial }{\partial
x}}\omega(x,t,z, p)\right )zt-$$$$-\left ({\frac {\partial
}{\partial x}}\omega(x,t,z,p) \right )\left ({\frac {\partial
}{\partial t}}\omega(x,t,z,p)\right )t -5\,\left ({\frac {\partial
}{\partial t}}\omega(x,t,z,p)\right ) \left ({\frac {\partial
}{\partial x}}\omega(x,t,z,p)\right ){\frac {
\partial }{\partial p}}\omega(x,t,z,p)+$$$$+p\alpha\,\left ({\frac {
\partial }{\partial p}}\omega(x,t,z,p)\right )^{2}+5\,x\left ({\frac {
\partial }{\partial z}}\omega(x,t,z,p)\right ){\frac {\partial }{
\partial p}}\omega(x,t,z,p)+z\alpha\,\left ({\frac {\partial }{
\partial z}}\omega(x,t,z,p)\right ){\frac {\partial }{\partial p}}
\omega(x,t,z,p)+$$$$+4\,t\left ({\frac {\partial }{\partial
t}}\omega(x,t,z ,p)\right )\alpha\,{\frac {\partial }{\partial
p}}\omega(x,t,z,p)+t \left ({\frac {\partial }{\partial
t}}\omega(x,t,z,p)\right )\nu\,{ \frac {\partial }{\partial
p}}\omega(x,t,z,p)-$$$$-x\nu\,\left ({\frac {
\partial }{\partial x}}\omega(x,t,z,p)\right ){\frac {\partial }{
\partial p}}\omega(x,t,z,p)+x\alpha\,\left ({\frac {\partial }{
\partial x}}\omega(x,t,z,p)\right ){\frac {\partial }{\partial p}}
\omega(x,t,z,p)=0. \end{equation}

     In spite of the fact that this equation looks not a simple than
equation (\ref{dryuma:eq25}) its  particular solutions can be find
without trouble.

    As example the substitution of the form
        \[
    \omega(x,t,z,p)=B(x,t,z)+k p t,
\]
into the equation (\ref{dryuma:eq26}) lead to expression on the
function $B(x,t,z)$
\[
B(x,t,z)=\alpha\,tx+tz+A(t),
\]
with arbitrary function $A(t)$ and the conditions on coefficients
of the R\"ossler system
\[
 \beta=1/5\,{\frac {\alpha\,\left
(8+5\,k\right )}{{k}^{2}}},\quad
 \nu={\frac {\alpha\,\left
(1+k\right )}{k}}
\]
depending from arbitrary parameter $k$.

    Using the function $\omega(x,t,z,p)$ we can
    obtain the solution of the equation (\ref{dryuma:eq23}) by
    elimination of the parameter $t$ from the relations
    \[
  \eta(x,y,z,p)-t{\frac {\partial }{\partial t}}\omega(x,t,z,p)+\omega(x
,t,z,p)=0,\quad y-{\frac {\partial }{\partial
t}}\omega(x,t,z,p)=0.
\]

As example at the choice $(A(t)=1+t^2$ we get
\[
\eta(x,y,z,p)=1/4\,{k}^{2}{p}^{2}+\left
(-1/2\,yk+1/2\,zk+1/2\,\alpha \,xk\right )p+1/4\,{z}^{2}+\left
(1/2\,\alpha\,x-1/2\,y\right )z-1+1/4
\,{y}^{2}-1/2\,y\alpha\,x+1/4\,{\alpha}^{2}{x}^{2}.
 \]

At the condition $A(t)=\ln(t)$ we get
\[
\eta(x,y,z,p)=1-\ln (-\left (-y+\alpha\,x+z+kp\right )^{-1}).
\]

    Remark that parameters $\alpha,\beta,\nu$
   in these cases the relation (\ref{dryuma:eq24}) is satisfied.

      To cite another example.

      Substitution the expression
\[
      \omega(x,t,z,p)=A(t)x+tz+C(t)p
\]
into the equation (\ref{dryuma:eq26}) lead to the system of
equations on the functions $A(t),~C(t)$
\[
\left (t\nu\,C(t)+4\,t\alpha\,C(t)-5\,A(t)C(t)-A(t)t\right ){\frac
{d} {dt}}A(t)+\left (A(t)\right
)^{2}-\nu\,A(t)C(t)+\alpha\,A(t)C(t)=0,
\]
\[
\left (t\nu\,C(t)+4\,t\alpha\,C(t)-5\,A(t)C(t)-A(t)t\right ){\frac
{d} {dt}}C(t)+A(t)C(t)-\nu\,\left (C(t)\right )^{2}+\alpha\,\left
(C(t) \right )^{2}=0,
\]
\[
-5\,\beta\,z\left (C(t)\right )^{2}+\left
(5\,z\alpha\,t-5\,A(t)z+5\,z \nu\,t\right )C(t)+3\,A(t)zt=0.
\]

     From this system of equations we find
     the condition on parameters
 \begin{equation}\label{dryuma:eq27}
     \nu=3/5\,\beta-\alpha
\end{equation}
and expression on the function $A(t)$
\[A(t)=-\beta\,C(t).
\]

    Function $C(t)$ in this case satisfies  the equation
     \[
     \left (25\,\beta\,C(t)+8\,\beta\,t+15\,t\alpha\right ){\frac {d}{dt}}C
(t)-8\,\beta\,C(t)+10\,\alpha\,C(t) =0.
\]

     Its  solution is defined by the relation
     \[
     -t\alpha+\left (C(t)\right )^{-1/2\,{\frac {8\,\beta+15\,\alpha}{-4\,
\beta+5\,\alpha}}}{\it \_C1}\,\alpha-\beta\,C(t) =0.
\]

     Using the expression on the function $C(t)$ we can find the function $\omega(x,t,z,p)$ and
     after elimination of the parameter $t$ from corresponding relations
     it is possible to get the solution of the
     eikonal equation at the condition (\ref{dryuma:eq27}) on parameters of the R\"ossler system of equations.

 \end{document}